
\def~{^}

\def\a{\alpha}

\def\d{\delta}

\def\ve{\varepsilon}
\def\z{\zeta}
\def\t{\theta}

\def\lb{\lambda}

\def\vp{\varphi}

\font\tenbb=msym10
\font\sevenbb=msym7
\font\fivebb=msym5
\newfam\bbfam
\textfont\bbfam=\tenbb \scriptfont\bbfam=\sevenbb
\scriptscriptfont\bbfam=\fivebb
\def\bb{\fam\bbfam}

\def\Cb{{\bb C}}

\def\Nb{{\bb N}}

\def\Rb{{\bb R}}

\def\Lc{{\cal L}}

\def\eqv{\equiv}

\def\ify{\infty}

\def\lgl{\langle}

\def\mpo{\mapsto}

\def\op{\oplus}

\def\ot{\otimes}

\def\ov{\overline}

\def\part{\partial}

\def\rgl{\rangle}

\def\sps{\supset}

\def\wdg{\wedge}

\def\wh{\widehat}

\def\wt{\widetilde}

\def\ra{\rightarrow}

\def\and{\mathop{\rm and}\nolimits}

\def\const{\mathop{\rm const}\nolimits}

\def\Hom{\mathop{\rm Hom}\nolimits}

\def\Int{\mathop{\rm Int}\nolimits}

\def\res{\mathop{\rm res}\nolimits}

\def\Vect{\mathop{\rm Vect}\nolimits}

\font\tenfm=eufm10

\catcode`\@=11
\def\displaylinesno #1{\displ@y\halign{
\hbox to\displaywidth{$\@lign\hfil\displaystyle##\hfil$}&
\llap{$##$}\crcr#1\crcr}}

\def\ldisplaylinesno #1{\displ@y\halign{
\hbox to\displaywidth{$\@lign\hfil\displaystyle##\hfil$}&
\kern-\displaywidth\rlap{$##$}
\tabskip\displaywidth\crcr#1\crcr}}
\catcode`\@=12

\def\buildrel#1\over#2{\mathrel{
\mathop{\kern 0pt#2}\limits~{#1}}}

\def\build#1_#2~#3{\mathrel{
\mathop{\kern 0pt#1}\limits_{#2}~{#3}}}

\def\hfl#1#2{\smash{\mathop{\hbox to 6mm{\rightarrowfill}}
\limits~{\scriptstyle#1}_{\scriptstyle#2}}}

\def\hfll#1#2{\smash{\mathop{\hbox to 6mm{\leftarrowfill}}
\limits~{\scriptstyle#1}_{\scriptstyle#2}}}

\def\up#1{\raise 1ex\hbox{\sevenrm#1}}

\def\cqfd{\unskip\kern 6pt\penalty 500
\raise -2pt\hbox{\vrule\vbox to10pt{\hrule width 4pt
\vfill\hrule}\vrule}\par}

\def\signed#1 (#2){{\unskip\nobreak\hfil\penalty 50
\hskip 2em\null\nobreak\hfil\sl#1\/ \rm(#2)
\parfillskip=0pt\finalhyphendemerits=0\par}}

\newfam\bffam \textfont\bffam=\tenbf \scriptfont\bffam=\sevenbf
\scriptscriptfont\bffam=\fivebf
\def\bf{\fam\bffam\tenbf}

\def\cc#1{\hfill\kern .7em#1\kern .7em\hfill}

\catcode`\@=11
\def\system#1{\left\{\null\,\vcenter{\openup1\jot\m@th
\ialign{\strut\hfil$##$&$##$\hfil&&\enspace$##$\enspace&
\hfil$##$&$##$\hfil\crcr#1\crcr}}\right.}
\catcode`\@=12

\def\boxit#1#2{\setbox1=\hbox{\kern#1{#2}\kern#1}%
\dimen1=\ht1 \advance\dimen1 by #1 \dimen2=\dp1 \advance\dimen2 by #1
\setbox1=\hbox{\vrule height\dimen1 depth\dimen2\box1\vrule}%
\setbox1=\vbox{\hrule\box1\hrule}%
\advance\dimen1 by .4pt \ht1=\dimen1
\advance\dimen2 by .4pt \dp1=\dimen2 \box1\relax}

\font\twelverm=cmr12

\magnification=1200

\overfullrule=0pt

\vglue 1cm

\centerline{\twelverm Poisson--Lie group of pseudodifferential symbols}

\smallskip

\centerline{\twelverm and fractional KP-KdV hierarchies}

\vglue 1cm

\centerline{Boris KHESIN}

\centerline{I.H.E.S., 35 route de Chartres, 91440 Bures-sur-Yvette, FRANCE}

\centerline{Department of Mathematics, Yale University, New Haven, CT 06520,
USA}

\medskip

\centerline{and}

\medskip

\centerline{Ilya ZAKHAREVICH}

\centerline{Department of Mathematics, M.I.T., Cambridge, MA 02139, USA}

\vglue 4cm

\noindent {\bf Abstract.} The Lie algebra of pseudodifferential symbols on the
circle has a nontrivial central extension (by the ``logarithmic'' 2-cocycle)
generalizing the Virasoro algebra. The corresponding extended subalgebra
of integral operators generates the Lie group of classical symbols of all
real (or complex) degrees. It turns out
that this group has a natural Poisson-Lie structure whose restriction to
differential operators of an arbitrary integer order coincides with the second
Adler-Gelfand-Dickey structure. Moreover, for
any real (or complex) $\a$ there exists a
hierarchy of completely integrable equations  on
 the degree $\a$ pseudodifferential symbols, and
this hierarchy for ${\a}=1$ coincides
 with the KP one, and for an integer ${\a}=n\geq 2$
and purely differential symbol gives the $n$-KdV-hierarchy.

\vfill\eject

\noindent {\bf 1.} We begin with a description of our main object,
the ring {\tenfm G} of
pseudodifferential symbols and of its central extension.

\medskip

This ring consists of formal series $A(x,D)=\build \sum_{-\ify}~{n}
a_i (x) D~i$ with respect to $D$ ($=d/dx$)
where $a_i \in C~{\ify} (S~1 , \Rb$ or $\Cb )$. The
multiplication law in {\tenfm G} is given by the Leibnitz
rule for multiplication of symbols: $A(x,\xi )\circ B(x,\xi )=\build
\sum_{n\geq 0}~{} {1\over n!} \ A_{\xi}~{(n)} (x,\xi ) \ B_x~{(n)} (x,\xi )$
where $A_{\xi}~{(n)} =d~n /d\xi~n A(x,\xi )$, $B_x~{(n)} =d~n /dx~n B(x,\xi
)$, and the Lie algebra structure on {\tenfm G} is natural: $[A,B]=A\circ B
-B\circ A$.
Recall also that the operator $\res :\hbox{\tenfm G} \ra C~{\ify} (S~1)$ is
defined by $\res \left( \sum a_i (x)D~i \right) =a_{-1} (x)$.

\medskip

Consider the formal expression $\log D$.
Certainly, $\log D \not \in $ {\tenfm G}, but for any
$A \in$ {\tenfm G} the formal commutator $[\log
D,A]=\log D\circ A-A\circ \log D$ is an element of {\tenfm G}. Thus $\log D$
acts on {\tenfm G} by commutation $[\log D,*]$, and explicitly this action is
given by: $[\log D,A] =$ \break
$ \build \sum_{k\geq 1}~{} {(-1)~{k+1} \over k} \
A_x~{(k)} D~{-k}$.

\bigskip

\noindent {\bf Theorem 1.} [10] The following 2-cocycle

$$c(A,B)=\int \res \left( [\log D,A]\circ B\right) =\int \res \left(
\sum_{k\geq 1}~{} {(-1)~{k+1} \over k} \ A_x~{(k)} D~{-k} \circ B \right)
\eqno (1)$$

\noindent gives a nontrivial central extension of the Lie algebra {\tenfm G}
(here $A$ and $B$ are arbitrary pseudodifferential symbols on
$\Rb$ or $S~1$). The
restrictions of this cocycle to the subalgebra $\hbox{\tenfm G}_{DO}$ of
differential operators and to the Lie algebra
of vector fields $\Vect $ are nontrivial.

\medskip

For $\hbox{\tenfm G}_{DO}$ the restriction gives the Kac-Peterson
cocycle [9],
for $\Vect (S~1)$ we get the Gelfand-Fuchs cocycle  which
defines the Virasoro algebra.

\medskip

\noindent {\bf Remark.} The other 2-cocycle on $\psi DS(S^1)$ is given by
$c'(A,B)=\int \res \left( [x,A]\circ B\right)$. One can define
the universal extension of $\psi DS(S^1)$ keeping the symmetry
of $x$ and $D$.

\vglue 1cm

\noindent {\bf 2.} Let $\wt{\hbox{\tenfm G}}$ be the ``double extension'' of
{\tenfm G}, i.e. we extend the algebra {\tenfm G} by the 2-cocycle $c(L,M)$ and
by the symbol $\log D:\wt{\hbox{\tenfm G}} =\left\{ \left( \build
\sum_{j=-\ify}~{n} a_j (x) D~j +\lb \log D,c\right)\right\}$. Here $c$ lies in
the center, the commutators of the rest are given by the Leibnitz rule above,
${\lb}\in \Rb \ {\rm or} \ \Cb$
and, finally, $c(\log D,M)=0$ for $M\in \hbox{\tenfm G}$.

\medskip

The algebra $\wt{\hbox{\tenfm G}}$ (as well as {\tenfm G}, cf.[7]) has an
$ad$-invariant nondegenerate inner product (``Killing form''):
$\left((A+\lb \log D,c) \ , \ (B+\mu \log D,d)\right) = 2\int \res (A\circ
B)+\lb d+\mu c$  for $A,B\in \hbox{\tenfm G}$
(the $ad$-invariance of this form is an
immediate corollary of the definition of $\wt{\hbox{\tenfm G}}$).
Moreover, this  algebra  has two remarkable subalgebras:
  1) $\wt{\hbox{\tenfm G}}_{DO}$ which is the algebra of centrally
extended differential operators $( \build \sum_{j\geq 0}~{} a_j (x) D~j
,c)$, where $c$ is the two cocycle
$c(A,B)=\int \res \left([\log D,A]\circ
B\right)$ and
 2) $\wt{\hbox{\tenfm G}}_{\Int}$ which is the algebra of integral
symbols together with $\log D : \{ \build \sum_{j=-\ify}~{-1}
a_j (x) D~j +\lb \log D\}$.

\bigskip

\noindent {\bf Theorem 2.} $\left( \wt{\hbox{\tenfm G}} , \wt{\hbox{\tenfm
G}}_{DO} , \wt{\hbox{\tenfm G}}_{\Int} \right)$ is a Manin triple (or,
equivalently, $\wt{\hbox{\tenfm G}}_{\Int}$ is a Lie bialgebra).

\medskip

The proof consists of the observation that $\wt{\hbox{\tenfm G}} =
\wt{\hbox{\tenfm G}}_{DO} \op \wt{\hbox{\tenfm G}}_{\Int}$ and the subalgebras
$\wt{\hbox{\tenfm G}}_- =\wt{\hbox{\tenfm G}}_{\Int}$ and
$\wt{\hbox{\tenfm G}}_+ =\wt{\hbox{\tenfm G}}_{DO}$ are isotropic with
respect to the inner product discussed above.

\bigskip

\noindent {\bf Corollary.} The Lie group $\wt{\hbox{\tenfm G}}_{\Int}$
corresponding to the Lie bialgebra $\wh{\hbox{\tenfm G}}_{\Int}$ has a natural
Poisson-Lie structure (the next section contains a detailed description of this
group).

\medskip

This structure can be defined by the following bivector $r\in
\wt{\hbox{\tenfm G}} \wdg \wt{\hbox{\tenfm G}} :\lgl r,\ov{a}_+ \wdg \ov{a}_-
\rgl = - \lgl r, \ov{a}_- \wdg \ov{a}_+ \rgl =(a_+ ,a_- )$. Here $a_{\pm} \in
\wt{\hbox{\tenfm G}}_{\pm}$, by $\ov a$ we denote the element of
$\wt{\hbox{\tenfm G}}~*$ dual to $a\in \wt{\hbox{\tenfm G}} :\lgl \ov a ,*\rgl
=(a,*)$, where $\lgl *,*\rgl$ is the natural pairing and $(*,*)$ is the
Killing form. If we identify the space $\wt{\hbox{\tenfm G}} \ot
\wt{\hbox{\tenfm G}}$ with $\Hom (\wt{\hbox{\tenfm G}} ,\wt{\hbox{\tenfm G}})$
using the pairing on {\tenfm G} then the bivector $r$
 corresponds to the skew symmetric
operator $\ov r$ such that
$\ov r |_{\wt{\hbox{\tenfm G}}_-} =-1$, $\ov r |_{\wt{\hbox{\tenfm G}}_+} =1$.

\medskip

To recall following [3],[12],[8] explicit formulae for the group Lie-Poisson
structure let $\wt G$, $\wt{G}_+$ and $\wt{G}_-$ be the Lie groups
corresponding to the Lie algebras {\tenfm G}, $\wt{\hbox{\tenfm G}}_+$ and
$\wt{\hbox{\tenfm G}}_-$, and let $\t$ and $\z$ be cotangent vectors to
$\wt{G}_-$ at a point $g\in \wt{G}_- \ (\t ,\z \in T_g~* \ \wt{G}_- )$. Extend
them arbitrarily to cotangent vectors (at $g$) to the larger group $\wt G \sps
\wt{G}_-$ (we denote these extensions by the letters $\t'$ and $\z'$).

\bigskip

\noindent {\bf Proposition.} [3],[12],[8] The Poisson structure on $\wt{G}_-$
is
defined by the following formula

$$\pi_{G_-} (\t ,\z )=\left( (\t')_+ ,\z' \right)
 -\left( ( Ad_g~* \ \t' )_+ ,Ad_g~*
\ \z' \right) .$$

\vglue 1cm

\noindent {\bf 3.} In this section we describe the structure and geometry of
the Lie group $\wt{G}_{\Int}$ cor\-responding to the Lie algebra
$\wt{\hbox{\tenfm G}}_{\Int}$ of integral symbols extended by the $\log D$.

\medskip

While the Lie algebra $\wt{\hbox{\tenfm G}}_{DO}$ of differential operators
does not seem to have a \break reasonable Lie group,
 the dual part of the Manin triple, i.e. the
algebra  $\wt{\hbox{\tenfm G}}_{\Int}$ admits the Lie group with the following
simple description.

\bigskip

\noindent {\bf Definition.} Classical Volterra's pseudodifferential symbol
$(\psi DS)$ is an expression of the form $P=\left( 1+\build \sum_{k=-\ify}~{-1}
u_k (x) D~k \right) \circ D~{\a}$, where ${\a}\in \Rb$ or $\Cb$, $D=d/dx$, $u_k
\in C~{\ify}$ ($\Rb$ or $S~1$). Call the (real or complex) number $\a$ the
degree of the symbol $P$. The multiplication of the symbols is uniquely defined
by the commutation relation
$[D~{\a} , u(x)]=\build \sum_{\ell \geq 1}~{} \left(
{{\a}\atop \ell}\right) u~{(\ell )} (x) D~{{\a}-\ell}$ where $\left( {{\a}\atop
\ell}\right) ={{\a}({\a}-1)\cdots ({\a}-\ell +1) \over \ell !}$.
 All such symbols form a
group with respect to this product.

\medskip

Define on the set of $\psi DS$ the topology as the standard topology on
the line in
the direction of $\a$ and the topology of the projective limit along the
variable $k$. For an individual $k$ we consider a usual $C~{\ify} (\Rb)$ (or
$C~{\ify} (S~1 )$) topology on the coefficients $u_k$. Then the basic
neighborhoods of a point $P~{(0)}$ are the sets of $P$'s such
that $|{\a}-{\a}~{(0)}
|<\ve$, $|u_k (x)-u_k~{(0)} (x)|<\vp (x)$, $k=0,\ldots ,\ell$ for fixed $\ve$,
$\ell$ and a fixed positive function $\vp (x)$.

\bigskip

\noindent {\bf Theorem 3.} The set of classical Volterra's symbols equipped
with
this topology forms a Lie group (denote it as $\wt{G}_{\rm Int}$).
 The corresponding
 Lie algebra coincides
with $\wt{\hbox{\tenfm G}}_{\rm Int}$.

\bigskip

\noindent {\bf Remark.} It is evident that the Lie algebra for the subgroup
$\{ ( 1+\build \sum_{k=-\ify}~{-1} u_k D~k ) \}$ is
$\hbox{\tenfm G}_{\rm int}$ consisting of integral symbols $\build
\sum_{k=-\ify}~{-1} u_k D~k$. Heuristically,
the tangent vector to $D~{\a}$ can be
obtained by differentiation of
this 1-parameter subgroup with respect to $\a$ at
${\a}=0$:
${d/ d{\a}} \biggl|_{\a =0} \ D~{\a}
=\log D\circ D~{\a} \Bigl|_{\a =0} =\log D.$

 To define the exponential map $\wt{\hbox{\tenfm G}}_{\rm Int} \ra
\wt{G}_{\rm Int}$ let us fix an integral symbol $A$.
 The would be exponent $P~{(t)} = \exp (t(\lb \log D+A))$
 of the symbol $\lb \log D+A$ should
satisfy the equation
$\left( {d\over dt} \ P~{(t)} \right) \circ \left( P~{(t)} \right)~{-1}
=\lb \log D+A .$

\bigskip

\noindent {\bf Theorem 4.} The exponential map $\exp :\wt{\hbox{\tenfm G}}_{\rm
Int} \ra \wt{G}_{\rm Int}$ given by the relation \break
$t(\lb \log D+A) \mpo P~{(t)}$
 is well-defined on
the entire Lie algebra $\wt{\hbox{\tenfm G}}_{\rm Int}$ and for fixed $t\not =
0$ it is a bijection between $\wt{\hbox{\tenfm G}}_{\rm Int}$ and $\wt{G}_{\rm
Int}$.

\bigskip

\noindent {\bf Remark.} The group $\wt{G}_{\rm Int}$ is an  infinite
dimensional analog of a unipotent group. Analogously to the finite dimensional
case where the exponential map is one-to-one and the exponential series
consists of a
finite number of terms, in our situation for $\psi DS$ every coefficient in $P
=\exp (\log D+A)$ is defined by a finite number of terms of the symbol $A$.

\vglue 1cm

\noindent {\bf 4.} In this section we show how the general Poisson-Lie group
techniques (see [3, 12, 8]) can be applied to the group of pseudodifferential
symbols. As a corollary of these constructions we obtain the Gelfand-Dickey-
(also called Adler-Gelfand-Dickey-
 or generalized \break KdV-)structures ([1],[5]) on the symbols.

\bigskip

\noindent {\bf Definition.} The (second generalized) Gelfand-Dickey Poisson
structure  on $\wt{G}_{\rm Int} =$ \break $\{ ( 1+\build
\sum_{k=-\ify}~{-1} u_k (x)D~k ) \circ D~{\a} \}$ is defined
as follows:

\medskip

\noindent a) The value of the Poisson bracket of two functions at the given
point is determined by the restriction of these functions on the subset
${\a}=\const$.

\medskip

\noindent b) The
subset ${\a}=\const$ is an affine space, so we can identify the
tangent space to this subset with the set of operators of the form $\d L=(
\build \sum_{k=-\ify}~{-1} \d u_k D~k ) D~{\a}$. We can also identify the
cotangent space with the space of operators of the form $X=D~{-\a} \ \ov X$,
where $\ov X$ is a differential operator, using the pairing
$F_X (\d L) \build =_{}~{\rm def} \lgl X,\d L\rgl =\int \res \d L\circ X .$

\medskip

\noindent c) Now it is sufficient to define the bracket on linear functionals,
and \break $\left\{ F_X ,F_Y \right\} \Bigl|_L =F_Y (V_{F_X} (L)),$
  where $V_{F_X}$ is the following Hamiltonian mapping \break
$F_X \mpo
V_{F_X} (L)$ (from cotangent space $D~{-\a} \circ \wt{\hbox{\tenfm G}}_{DO}$ to
the tangent space $\wt{\hbox{\tenfm G}}_{\Int} \circ D~\a$):
\break $V_{F_X} (L)=(LX)_+ \ L-L(XL)_+ \ .$

\bigskip

\noindent {\bf Remark.} Usually this definition is given only in the case when
$\a$ is a fixed positive integer and $L$ is a differential operator, cf.[1],
[5], [7].

\medskip

\noindent {\bf Theorem 5.} The Poisson
structure $\pi_{\wt{G}_{\rm Int}}$ (described in section 2) on the Poisson-Lie
group $\wt{G}_{\rm Int}$ coincides with the second generalized Gelfand-Dickey
structure.

\bigskip

\noindent {\bf Corollary.} The function ${\a}=\deg L$ of degree of $\psi DS$ is
the Casimir function, i.e. $\{ \deg L,\vp (L)\} =0$ for any function $\vp (L)$
on $\wt{G}_{\rm Int}$. Hence we can restrict the bracket to any level $\deg
L=\const$.

\medskip

\noindent {\bf Remark.} This is an analogue of the $GL_n$-Gelfand-Dickey
structure. The analog of the $SL_n$-structure on the submanifold $u_{-1}(x)\eqv
0$ is a result of the Poisson reduction along the action on   $\wt{G}_{\rm
Int}$
of the multiplicative group of nonvanishing functions.

\bigskip

\noindent {\bf Remark.} This viewpoint can be useful
 to describe the relation of the
$W_n$-  to $W_{\infty}$-algebras appeared recently in theoretical physics
(see e.g.[11]).
The classical $W_n$-algebras (or $SL_n$-Gelfand-Dickey algebras) are the
Poisson algebras of functions on the sections of the subgroup of
 $\wt{G}_{\rm Int}$ (for which $\{ u_{-1} (x) \eqv 0\}$)
 by hyperplanes $\a = n$. Let us  linearize this quadratic Poisson structure at
the identity element of
$\wt{G}_{\rm Int}$. This linear Poisson structure on
$\wt{\hbox{\tenfm G}}_{\rm Int}$ is exactly
the Lie algebra structure on $\wt{\hbox{\tenfm G}}_{\rm DO}$, i.e. on
 the centrally extended Lie algebra
of all differential operators, and it has been recognized as $W_{1+\infty}$
([2]).
The condition $u_{-1} (x) \eqv 0$ implies discarding of functions,
i.e. of zero order differential operators in
$\wt{\hbox{\tenfm G}}_{\rm DO}$
and gives exactly the $W_{\infty}$-algebra.

\vglue 1cm

\noindent {\bf 5.} Finally, we describe  integrable hierarchies
related to the construction discussed.

\medskip

Let us fix ${\a}\not = 0$. The corresponding ``hyperplane'' $\Lc_{\a} =$
\break $\{ L=(
1+\build \sum_{k=-\ify}~{-1} u_k D~k ) \circ D~{\a}
\}$ in $\wt{G}_{\rm Int}$ is a Poisson submanifold ($\a$ is a Casimir
functional). Consider the following infinite sequence of evolution equations on
the coefficients of $L\in \Lc_{\a}$:

$${\part L \over \part t_m} =\left[ L,(L~{m/{\a}})_+ \right] \ , \ m=1,2,\ldots
.\eqno (2)$$

\noindent Here the derivative ${\part \over \part t_m}$
denotes $m~{\rm th}$ flow
of the hierarchy. An arbitrary power of any operator in $\wt{G}_{\Int}$ is
well-defined due to the ``good properties'' of the exponential map
$\wt{\hbox{\tenfm G}}_{\Int} \ra \wt{G}_{\Int}$ (theorem 4). Indeed, the
operator $L$ defines uniquely the 1-parameter subgroup passing through it, and
the power of $L$ is the parameter along this subgroup.

\medskip

In particular, the degree of the operator $L~{m/{\a}}$ is
equal to $m$, i.e. to an
integer, and thus the operation $+$ (taking the differential part) does make
sense.

\bigskip

\noindent {\bf Theorem 6.} 1. For any positive integer $m$ the equation $(2)$
defines an evolution
on $\Lc_{\a}$ (i.e. the vector $\left[L,(L~{m/{\a}})_+ \right]$ is
tangent to $\Lc_{\a}$).

\medskip

\noindent 2. For any $m$ the equation is Hamiltonian on $\Lc_{\a}$ with
respect to
the (second generalized) Gelfand-Dickey bracket and the Hamiltonian function
$H_m (L)= {\a \over m} \int \res (L~{m/{\a}})$.

\medskip

\noindent 3. The sequence $\{ H_m \}$, $m\in \Nb$ is an infinite set of
integrals in involution with respect to the second Gelfand-Dickey structure.

\medskip

\noindent 3'. Any two flows of the hierarchy $(2)$ commute.

\bigskip

\noindent {\bf Remark.} For ${\a}=1$ we get nothing else but the standard
KP-hierarchy on $\Lc_1 =\{
(D+u_{-1}(x) D~0 +u_{-2}(x) D~{-1} +\cdots )\}$ (usually
to get the SL-case instead of the GL-one we put $u_{-1} \eqv 0$). For integer
${\a}=n\geq 2$ we  get the $n~{\rm th}$ KdV-hierarchy. Usually the
 classical KdV-flows are  considered only on the differential
part of the $\Lc_n$ (such a  Poisson submanifold in $\Lc_{\a}$ exists for
integer $\a$ only) rather than on
the entire space $\Lc_n$. Thus we may restrict ourselves by $u_{-n-1} \eqv
u_{-n-2} \eqv \cdots \eqv 0$.

\medskip

Thus within the framework of the Poisson-Lie group of $\psi DS$ it is natural
to define KP-KdV-hierarchies for any real (or complex) $\a$ (let us call them
fractional- or $\a$-KP-KdV-hierarchies).
This construction can be viewed as a natural generalization of
the approach of Gelfand and Dickey [4] to the $n~{\rm th}$ root of
an integer
 degree $n$ differential operator.
 It would be interesting to find a relation of our construction of
integrable hierarchies to that in [6]. Proofs and applications of the
Poisson-Lie structure to the Poisson properties of the Miura transform and of
the dressing action on $\psi DS$ will appear elsewhere.

\vglue 1cm

\noindent {\bf Acknowledgments.} We are pleased to thank V. Arnold, J.
Bernstein, M. Gromov, \break S. Khoroshkin, T. Khovanova,
N. Reshetikhin, C. Roger, M.
Semenov-Tian-Shansky and A. Weinstein for fruitful discussions.
We are indebted to B. Enriquez  for his help with French translation
of the short version of this paper.
 B. Khesin expresses his gratitude to
the Institut des Hautes Etudes Scientifiques in
Bures-sur-Yvette for its hospitality during the final stage of this work.

\vglue 2cm

\centerline{\twelverm REFERENCES}

\vglue 1cm

\smallskip

\item{1.} M. Adler, Inven. Math., Vol.50, 1979, pp.219-248.

\smallskip

\item{2.} I. Bakas, B. Khesin and E. Kiritsis : The logarithm of the
derivative operator and higher spin algebras of $W_{\infty}$ type.
Preprint UCB-PTH-91/48 , 1991,  Comm. Math. Phys.,1992 (to appear).

\smallskip

\item{3.} V. Drinfeld, Sov. Math. Doklady,
Vol.27,1983, pp.68-71.

\smallskip

\item{4.} I.M.Gelfand and L.A.Dickey, Funct. Anal. Appl., Vol.10, 1976,
pp.259-273.

\smallskip

\item{5.} I.M.Gelfand and L.A.Dickey, A family of Hamiltonian structures
associated with
nonlinear integrable differential equations. Preprint IPM AN SSSR - 139
1978.

\smallskip

\item{6.} M.F. de Groot, T.J. Hollowood and  J.L. Miramontes, Comm. Math.
Physics, Vol.145, n\up{0}1
1992, pp.57-84.

\smallskip

\item{7.} D.Lebedev and Y.I.Manin , Funct. Anal. Appl.,
Vol.13,1979, No.4, pp. 40-46.

\smallskip

\item{8.} J.H. Lu and A. Weinstein, J. of Diff. Geometry, Vol.31, 1990,
pp.501-526.

\smallskip

\item{9.} V.G. Kac and D.H. Peterson,  Proc. Nat. Acad. Sci. USA, Vol.78
1981, pp.3308-3312.

\smallskip

\item{10.} O.S. Kravchenko and B.A. Khesin, Funct. Anal. Appl., Vol.25,1991,
n\up{0}2, pp.83-85.

\smallskip

\item{11.} C.N. Pope, L.J. Romans and X. Shen, A brief history of $W_{\infty}$.
Preprint CTP TAMU-89-90, USC-90/HEP28, 1990.

\smallskip

\item{12.} M.A. Semenov-Tian-Shansky, Publ. RIMS, Kyoto Univ., Vol.21,1985,
pp.1237-1260.

\end